\documentclass[11pt]{article}
\usepackage{amsmath}
\usepackage{graphicx}
\usepackage{ textcomp }
\usepackage{ amssymb }
\usepackage{float}
\graphicspath{ {images/} }
\usepackage{booktabs} 
\usepackage[table]{xcolor} 
\usepackage{authblk}

\usepackage{tabularx} 
\usepackage{rotating} 
\usepackage{siunitx} 
\usepackage{caption}

\usepackage{hyperref}
\usepackage[numbers,sort&compress]{natbib}

\usepackage{adjustbox}
\usepackage[margin=0.5in]{geometry}
\begin{document}

\makeatletter
\newcommand*{\rom}[1]{\expandafter\@slowromancap\romannumeral #1@}
\makeatother

\title{High fitness paths can connect proteins with low sequence overlap}

\author{
  Pranav Kantroo\textsuperscript{1,6,}\footnote{pranav.kantroo@yale.edu} ,
  G\"unter P. Wagner\textsuperscript{2,3,4}, 
  Benjamin B. Machta\textsuperscript{5,6,}\footnote{benjamin.machta@yale.edu}
  \\
  \small
  \textsuperscript{1}Computational Biology and Bioinformatics Program, Yale University, New Haven, CT-06520, USA \\
  \textsuperscript{2}Emeritus, Department of Ecology and Evolutionary Biology, Yale University, New Haven, CT-06520, USA \\
  \textsuperscript{3}Department of Evolutionary Biology, University of Vienna, Djerassi Platz 1, A-1030 Vienna, Austria \\
  \textsuperscript{4}Hagler Institute for Advanced Studies, Texas A\&M, College Station, TX-77843, USA \\
  \textsuperscript{5}Department of Physics, Yale University, New Haven, CT-06520, USA \\
  \textsuperscript{6}Quantitative Biology Institute, Yale University, New Haven, CT-06520, USA 

}
\date{}
\maketitle
\vspace{-3em}

\begin{abstract}
\noindent The structure and function of a protein are determined by its amino acid sequence. While random mutations change a protein's sequence, evolutionary forces shape its structural fold and biological activity. Studies have shown that neutral networks can connect a local region of sequence space by single residue mutations that preserve viability. However, the larger-scale connectedness of protein morphospace remains poorly understood. Recent advances in artificial intelligence have enabled us to computationally predict a protein's structure and quantify its functional plausibility. Here we build on these tools to develop an algorithm that generates viable paths between distantly related extant protein pairs. The intermediate sequences in these paths differ by single residue changes over subsequent steps -- substitutions, insertions and deletions are admissible moves. Their fitness is evaluated using the protein language model ESM2, and maintained as high as possible subject to the constraints of the traversal. We document the qualitative variation across paths generated between progressively divergent protein pairs, some of which do not even acquire the same structural fold. The ease of interpolating between two sequences could be used as a proxy for the likelihood of homology between them.

\end{abstract}

\section{Introduction}

\begin{figure}[htbp]
    \centering
    \begin{minipage}{0.55\textwidth} 
        \centering
        \includegraphics[width=\textwidth]{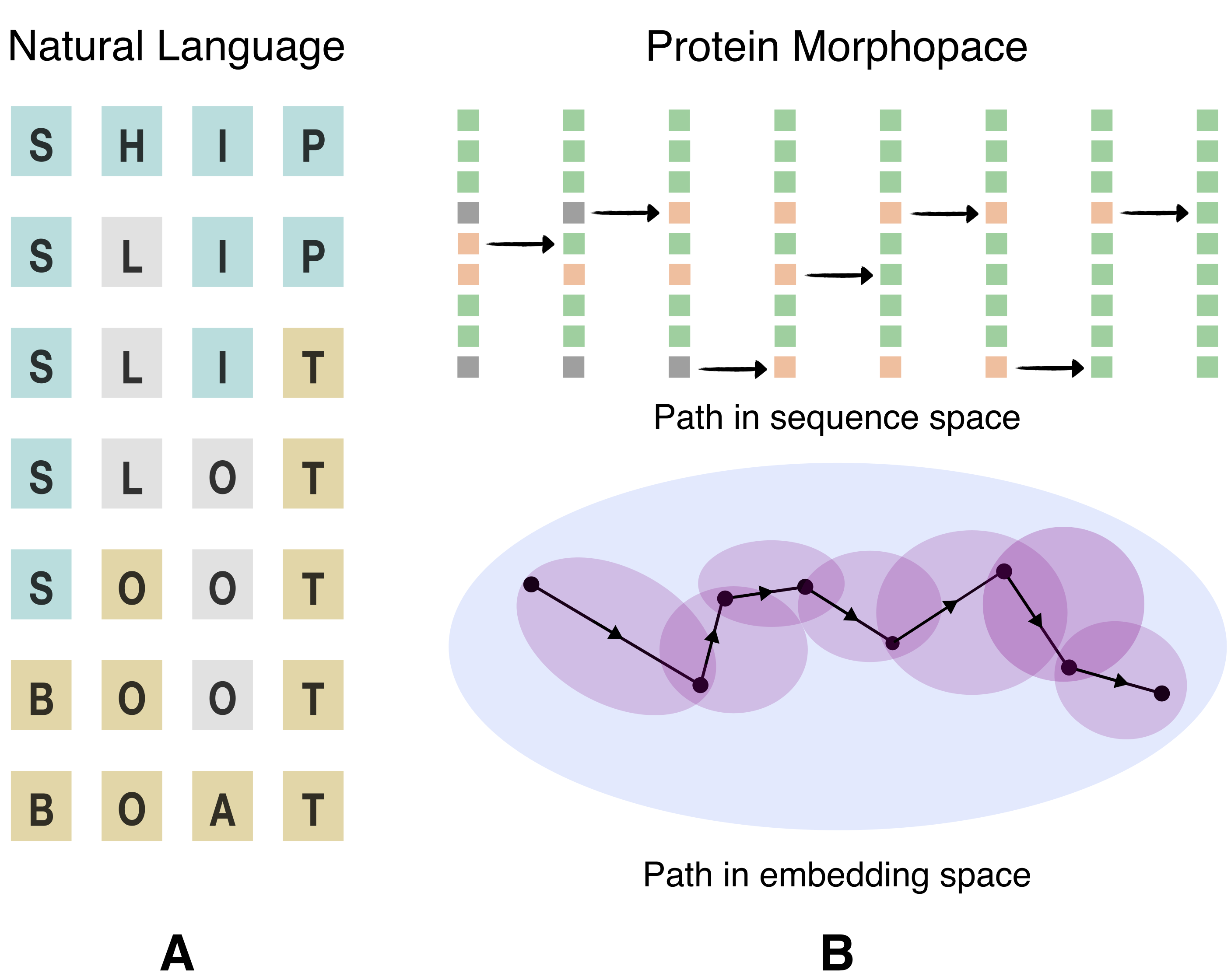}
    \end{minipage}
    \hfill
    \begin{minipage}{0.4\textwidth} 
        \captionof{figure}{\small \textbf{Maynard Smith's game} (A) The transformation from SHIP to BOAT as played in natural language. Each of the intermediates is a meaningful word in English, and subsequent words differ by a single substituted letter. 
                    (B) In place of the requirement that words be meaningful in English, we seek paths between amino acid sequences along which intermediates have a high fitness according to the protein language model ESM2. We show a schematic representation of a traversal through protein morphospace between two extant proteins. The intermediate sequences in such a traversal need not exist in nature. The changes in the source protein have been visualized in sequence space, as well as the embedding space of the protein language model. Single letter substitutions, insertions and deletions are admissible moves in this setting.}
        \label{fig:schematic}
    \end{minipage}
\end{figure}

Proteins catalyze metabolic reactions, facilitate signal transduction, act as structural elements in the cellular framework, and participate in myriad other biological processes essential to sustain life~\citep{alberts2022molecular}. While natural proteins are diverse, they constitute just a tiny fraction of possible amino acid sequences, most of which do not stably fold and may not be functional~\citep{keefe2001functional}. How does evolution navigate this space to find viable sequences? In a seminal work, Maynard Smith approached this question by invoking an analogy to a word game~\citep{maynard1970natural}. The goal of the game is to link two given words through single letter changes such that intermediate words remain meaningful~(Figure~\ref{fig:schematic}:A). In this analogy, the words are protein sequences, the single-letter changes that they undergo are mutations, and the requirement to preserve meaning translates to maintaining the functional viability of intermediate sequences. The empirical relevance of this picture is corroborated by the widespread prevalence of neutral networks -- a collection of sequences with virtually identical fitness that differ by single-residue mutations~\citep{manrubia2021genotypes, manrubia2010neutral, papkou2023rugged}. A population could presumably use this interconnected web of sequences to diffusively search for better adapted states~\citep{schultes2000one}. 

However, a quantitative assessment of the global connectivity structure of protein morphospace has remained elusive --  we have few tools to trace the possible mutational paths between divergent homologous proteins, especially so, when they differ in both sequence and structure. Here we present a computational scheme to play Maynard Smith's game with real protein sequences, where we judge the semantic coherence of the intermediates using assessments drawn from a protein language model~\citep{lin2022language} (Figure~\ref{fig:schematic}:B). 

\section{Results} 

The game begins with choosing a source and a target sequence. We use a stochastic variant of beam search to navigate through the mutational space that separates the two sequences while also maintaining the viability of the intermediates. This search protocol roughly entails sampling sensible mutations from some proposal distribution, and choosing functionally plausible variants within this set that are most similar to the target state over iterative steps. This search protocol relies on three core primitives: quantifying the sensibility of the move pool, assessing the viability of the candidate mutants, and ascertaining the distance between the candidates and the target state. We use the One Fell Swoop (OFS) approach~\citep{kantroo2024pseudo} to estimate both the substitution and the insertion profile of a sequence. These profiles are used as the proposal distribution to sample sensible candidate mutants, while deletions are sampled uniformly. We use the estimated substitution profile of a sequence to calculate its ESM2 OFS pseudo-perplexity, a scalar value that quantifies the functional plausibility of the input sequence: a lower value indicates a higher predicted fitness. We direct the paths to move towards the target state by using a sequence alignment-based proximity measure that also leverages ESM2 residue embeddings~\citep{pantolini2024embedding}. 

Below we present three examples of interpolating between endpoints that are chosen to represent three distinct levels of shared similarity: outright homologs with 36\% sequence identity that share the same structural fold, distant homologs with 18\% sequence identity that differ in some of their secondary structure motifs, and speculative homologs with 12\% sequence identity that acquire altogether different folds. We use paths generated through a random interpolation strategy as a control to assess the efficacy of the beam search protocol. The random strategy entails using the sequence alignment of the protein pair to identify non-matching positions in the two sequences. These non-matching positions for a chosen source are edited to match the target one-by-one in a randomized order until all the positions match. Paths generated via the random strategy thus always require as many steps as the hamming distance between the sequence alignments of the two proteins. 

\subsection{Outright Homologs}

\begin{figure}[h]
    \centering
    \includegraphics[width=1\textwidth]{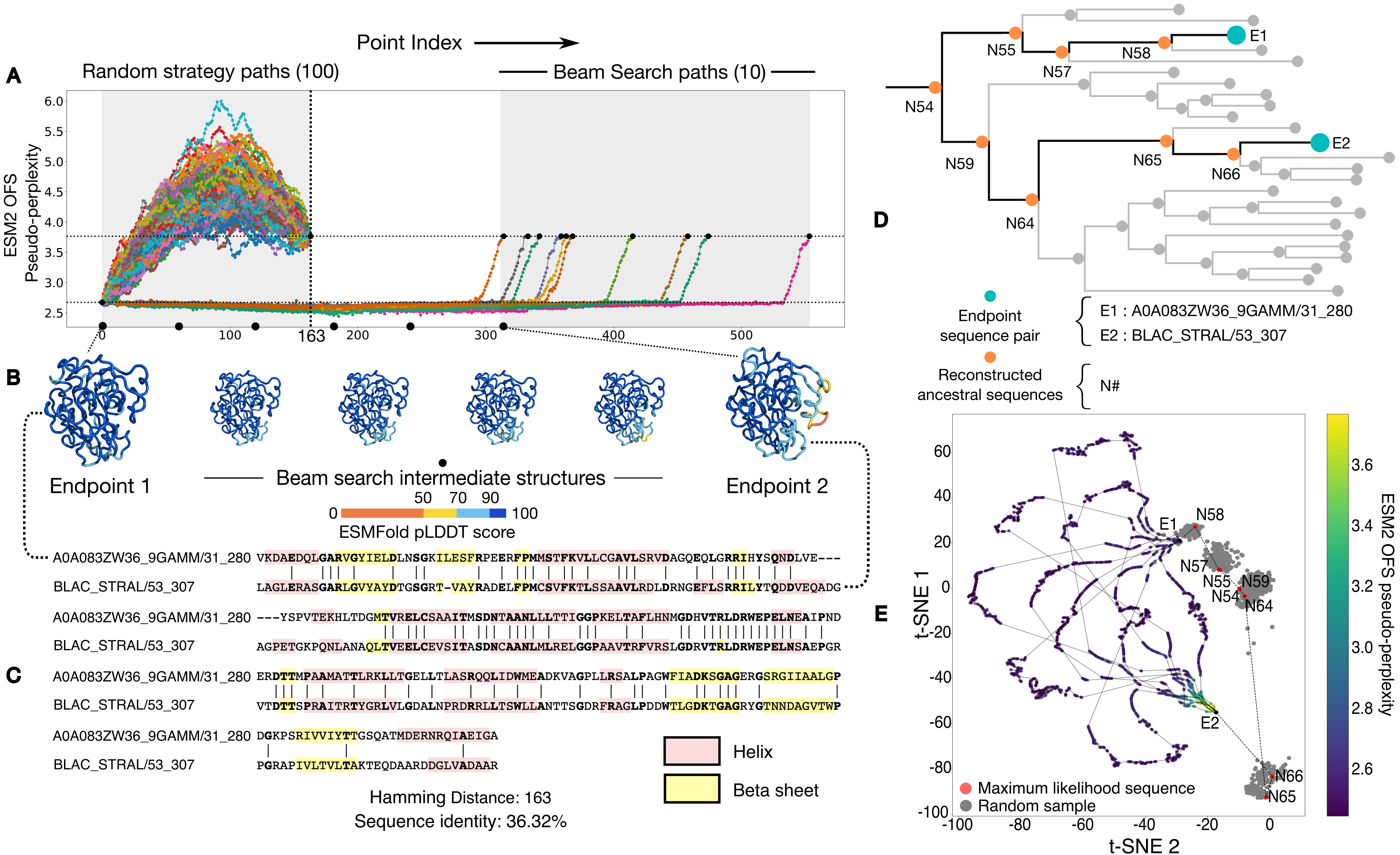}
    \caption{\small \textbf{Outright homologs can be connected by viable paths that preserve                           ESM2 OFS pseudoperplexity and maintain folded structure.} 
                            (A) We compare paths taken by the random strategy against paths generated by the beam search strategy. The random strategy paths pass through intermediates with a higher ESM2 OFS pseudo-perplexity than the endpoint on the right (dotted lines), while those generated by beam search do not. 
                            (B) A visualization of the folded structures of the intermediates along one of the trajectories. The folding confidence of the beam search intermediates remains high as seen by the ESMFold pLDDT scores. The beam search intermediates also maintain structural similarity with both of the endpoint structures.
                            (C) The sequence alignment between the two chosen sequences along with their secondary structure motifs.
                            (D) A visualization of the subtree within the phylogeny of the protein family that contains the endpoint sequence pair (blue) and their last common ancestor. We follow the path (orange) along the interior nodes in the subtree that connects the pair through their last common ancestor. 
                            (E) A t-SNE plot of the interpolated paths along with the inferred ancestral path determined using ancestral state reconstruction. The maximum likelihood sequences for the individual nodes are red, while other possible samples are colored in gray.}
    \label{fig:Outright}
\end{figure}

The $\beta$-lactamase family (Pfam ID: PF00144) of enzymes mediates resistance to $\beta$-lactam-based antibiotics in bacteria~\citep{tooke2019beta}. A landmark experiment~\citep{weinreich2006darwinian} used this family to assess the selective accessibility of mutational paths between two sequences that differ in 5 positions. The study considers 120 paths between the endpoints, created by making the 5 changes in all possible orders, and finds that only a small fraction of the paths monotonically increase fitness in response to antibiotic stress. Here we use a computational approximation for fitness, and rather than seeking a monotonically increasing value we merely look for approximate preservation of fitness along the entire path. This allows us to look for paths between endpoints that differ in significantly more positions, 163 rather than 5. (Figure~\ref{fig:Outright}:C).

We first generate 100 paths between the chosen pair using the random interpolation strategy. We find that the randomly generated paths pass through intermediates with a higher ESM2 OFS pseudo-perplexity than the endpoint sequences -- the high values are indicative of low fitness. We then generate 10 paths with our beam search protocol. In this case, we find that ESM2 OFS pseudo-perplexity of the intermediates remains in between the values of the endpoints. This indicates that the predicted functional plausibility of the beam search intermediates remains high throughout the traversal (Figure~\ref{fig:Outright}:A).

The outright homology between the pair means that we can use the phylogeny of the protein family to trace the inferred ancestral path between the source and the target through ancestral state reconstruction. We use the reconstruction process to sample the sequences corresponding to the interior nodes in the tree that connect the extant sequence pair through their last common ancestor (Figure~\ref{fig:Outright}:D). We visualize this inferred ancestral path along with the interpolated paths together on a t-SNE plot~\citep{van2008visualizing} by using the language model embeddings of the respective sequences for the projection. An analysis of the sequences of the intermediates in these paths reveals that the interpolated paths are different from each other and the inferred ancestral path. This in turn suggests that the local fitness landscape that the endpoint sequence pair is situated in, contains a dense network of functionally plausible sequences that can be connected by single residue mutations. 

\subsection{Distant Homologs}

\begin{figure}[!b]
    
    \centering
    \includegraphics[width=1\textwidth]{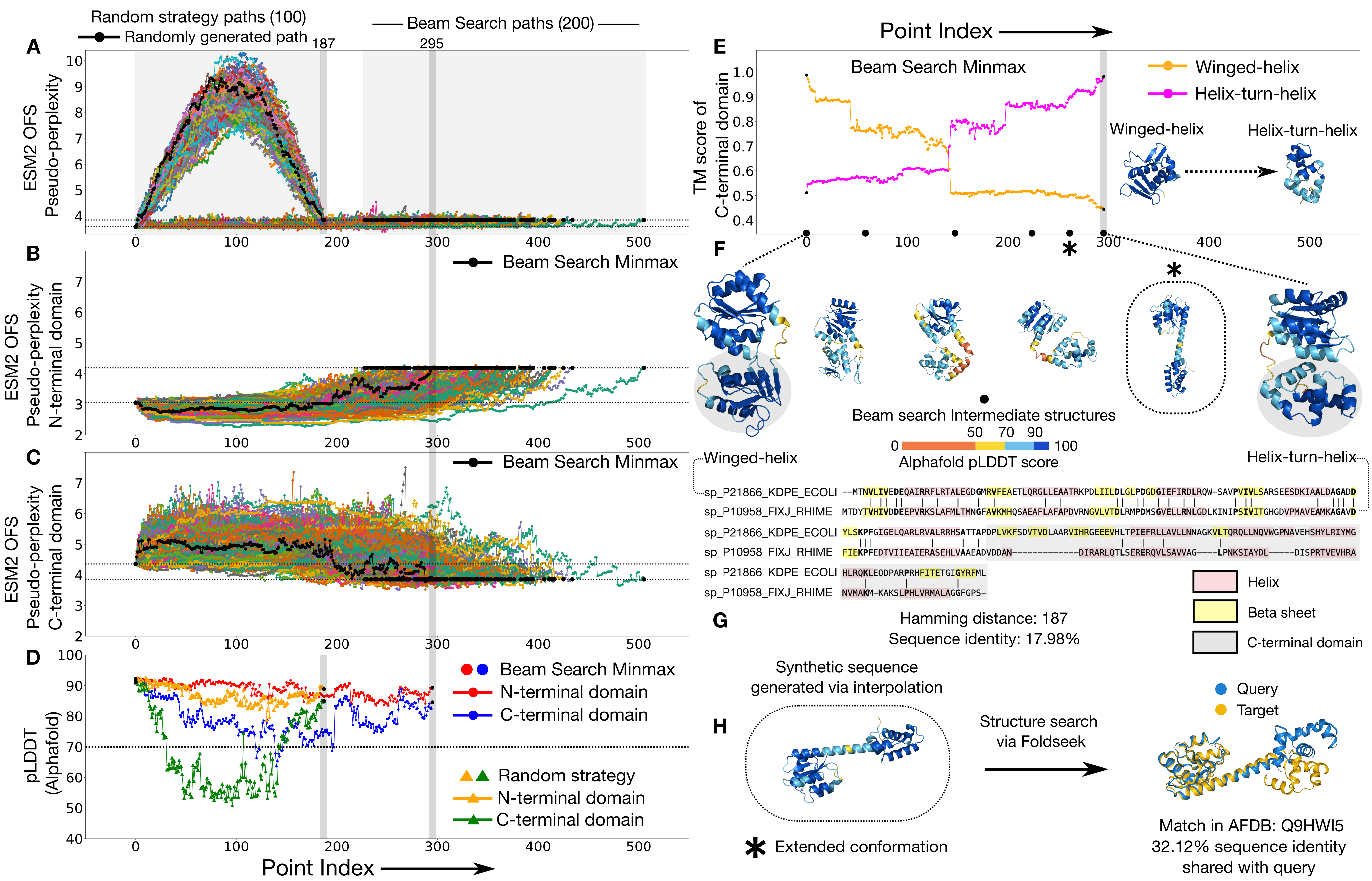}
    
    \caption{\small \textbf{Distant homologs with deviations in secondary structure can                    be connected by paths that visit structural conformations not shared                   by the endpoints.} 
                    (A) ESM2 OFS pseudo-perplexity of the beam search intermediates stays below the value of the endpoint on the right (dotted lines). The random strategy intermediates reach a much higher value, indicative of low fitness. The intermediates along one of the random strategy paths (black) were folded via Alphafold shown in D.
                    (B) The ESM2 OFS pseudo-perplexity of the N-terminal segment in the beam search intermediates stays below the endpoint on the right (dotted lines).
                    (C) The ESM2 OFS pseudo-perplexity of the C-terminal segment in the beam search intermediates is higher than the endpoint on the left (dotted lines). The path that minimizes the maximum ESM2 OFS pseudo-perplexity of the C-terminal domain is marked in black (also in B).
                    (D) The N-terminal segment of the minmax intermediates (red) consistently maintains a high folding confidence as measured by the pLDDT score, however there are some intermediates where the pLDDT score of the C-terminal segment (blue) dips below 70. The folding confidence for the random strategy intermediates tends to be lower than the minmax intermediates (orange and green).
                    (E) The structure of the C-terminal segment transitions from being more similar to the winged-helix domain to being more similar to the helix-turn-helix domain over the course of the path.
                    (F) One of the beam search intermediate structures acquires a distinctive extended conformation (\textasteriskcentered) that is not shared by either of the endpoints.
                    (G) The sequence alignment between the two endpoint sequences along with their secondary structure motifs.
                    (H) A structure-based search of this extended conformation reveals a match in the Alphafold database that shares low sequence identity with the synthetic sequence.}
    \label{fig:Distant}
\end{figure}

The response regulator protein family plays a role in effecting a bacterium's response to environmental cues~\citep{koretke2000evolution, gao2007bacterial, stock1989three, wright2018architecture, chakravarty2023identification}. These proteins are composed of a common N-terminal domain that functions as a phosphoacceptor, while the identity of their C-terminal output domain varies, and can take on various physiological roles involving signal transduction~\citep{galperin2010diversity}. It has been proposed that two of the output domains found in this family, the helix-turn-helix domain and the winged-helix domain, are related by descent~\citep{aravind2005many}. Both of these domains function as DNA binding domains, however they differ in sequence and some of their secondary structure motifs (Figure~\ref{fig:Distant}:G). A recent study argues that these two domains diverged via single residue changes through a phylogenetic analysis of the subfamilies~\citep{chakravarty2023identification}. This evolved fold-switching phenomenon~\citep{chakravarty2023identification} makes the families a natural choice to explicitly generate pointwise mutational paths and examine the changes in structure over the traversal.

We generate 100 paths (Uniprot IDs: P21866, P10958) using the random strategy and compare them against 200 paths generated through the beam search protocol (Figure \ref{fig:Distant}:A). We find that the beam search paths do significantly better than the randomly generated paths on the predicted fitness of the intermediates as measured by their ESM2 OFS pseudo-perplexity values (Figure \ref{fig:Distant}:A). Given the multi-domain nature of the protein, we also assessed the local fitness of the domains over the course of the transit. We found that the predicted fitness for the N-terminal domain remains high all the way through, for all of the beam search paths (Figure \ref{fig:Distant}:B). However, the paths pass through intermediates with a lower predicted fitness for the C-terminal domain in relation to the source and the target (Figure \ref{fig:Distant}:C). We folded the path that minimizes the maximum ESM2 OFS pseudo-perplexity for the C-terminal domain using Alphafold~\citep{jumper2021highly, mirdita2022colabfold}, and found a similar trend: the folding confidence as measured by the local pLDDT score for the N-terminal domain remains high, however it dips to lower values for the C-terminal domain. We also folded a path generated by the random strategy and found that the randomly generated intermediates tend to have a lower folding confidence as compared to the beam search intermediates~(Figure~\ref{fig:Distant}:D).

We examined the structures of the intermediates in the minmax folded path and identified a distinctive extended conformation (marked \textasteriskcentered\  in \ref{fig:Distant}:F) not shared by either of the endpoints (Figure~\ref{fig:Distant}:F). We ran a structure-based search of this extended conformation using FoldSeek~\citep{van2024fast} and found a match in the Alphafold Database that shares a low sequence identity (32.12\%) with the synthetic sequence (Figure~\ref{fig:Distant}:H). This implies that the procedure generated a sequence that acquires a distinctive structural conformation which is also found in naturally-occurring members of the family. This shows that paths generated by the search protocol can demonstrably pass through structural conformations found in nature that differ from the endpoint structures.

\subsection{Speculative Homologs}

\begin{figure}[h]
    \centering
    \includegraphics[width=1\textwidth]{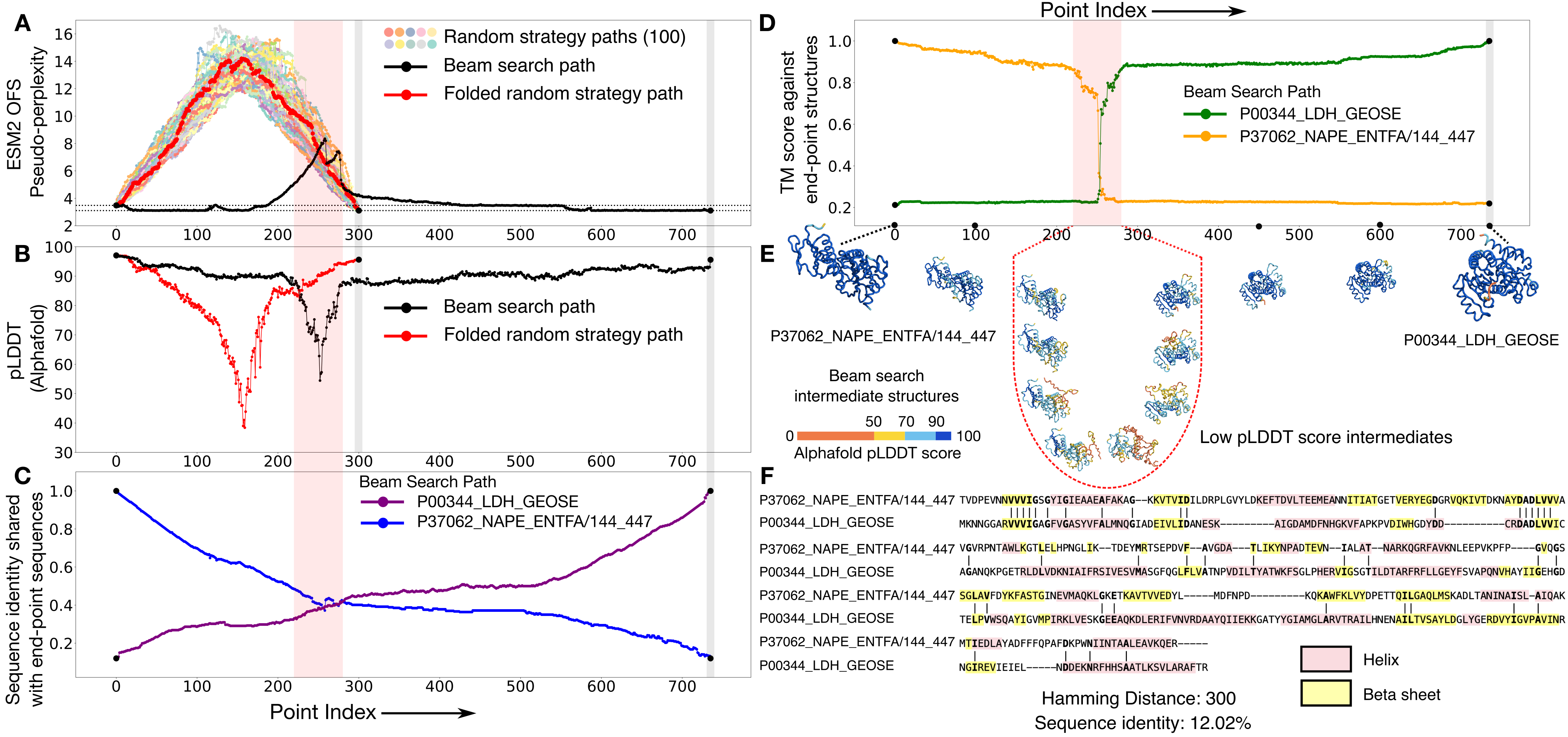}
    \caption{ \small \textbf{Speculative homologs that acquire altogether different                           structural folds can be connected by paths that                                  go through a brief span of high ESM2 OFS pseudo-perplexity values during which their fold switches                               abruptly.} 
                             (A) The paths generated by the random strategy (colors) perform especially poorly for this pair as seen by the high ESM2 OFS pseudo-perplexity of the intermediates. The representative beam search path (black) also passes through a stretch of low fitness intermediates. However, this stretch is shorter relative to the paths generated via the random strategy, and the maximal pseudo-perplexity value is also lower in comparison. One of the random strategy paths is chosen to be folded via Alphafold and is marked in red.
                             (B) The folding confidence of the beam search intermediates (black) as measured by the Alphafold pLDDT scores dips in the same span, as the stretch of low fitness scores (red bar). The folding confidence of the random strategy path (red) is also lower than the beam search path in comparison. 
                             (C) The sequence identity curve shows that some of the intermediates share a significant sequence identity with both of the endpoints simultaneously, which is in turn indicative of their chimeric character.
                             (D) Folding the intermediates along the beam search path shows that the structural transition is quite sharp between this pair, as seen by the sharp change in the TM score measured against the endpoint structures. 
                             (E) A visualization of the structural transition between the pair. The transition is mediated by structures that score low on folding confidence as seen by their pLDDT scores.
                             (F) The endpoint sequences share some similarity in their secondary structure motifs and also have two contiguous stretches of five and six identical residues: VVVIG and DADLVV}
    \label{fig:Speculative}
\end{figure}

The lactate dehydrogenase and the NADH peroxidase protein families both function as enzymes but differ in the specific metabolic reaction that they catalyze. Despite the significant differences in their three-dimensional structures, similarity in some of their secondary structure motifs has been used to make a case for possible shared homology between them~\citep{grishin2001fold}. There is also sequence identity observed in key positions~\citep{grishin2001fold} within some of the members: for the pair that we use (Uniprot ID: P00344, P37062), there are two contiguous stretches of five and six identical residues in the sequence alignment (Figure~\ref{fig:Speculative}:F). The limited overlap in their feature sets justifies the speculative homologs moniker.

We generate 100 paths between this pair using the random interpolation strategy, and compare them against one representative path generated through the beam search protocol (Figure~\ref{fig:Speculative}:A). ESM2 OFS pseudo-perplexity reaches fairly high values for the randomly generated paths, indicating that the randomized residue-replacement scheme performs especially poorly when the endpoints share little similarity. The beam search path also passes through such a fitness canyon, however the span of the low fitness values is comparatively brief, and the peak value of the pseudo-perplexity metric is not as high. We folded the intermediate sequences in this path using Alphafold~\citep{jumper2021highly, mirdita2022colabfold}, and found that the dip in the predicted fitness coincides with the lowered folding confidence of the intermediates as measured by their pLDDT scores (Figure~\ref{fig:Speculative}:B and Figure~\ref{fig:Speculative}:E). Comparing the folded intermediates against the endpoint structures reveals that this region also marks the cross-over point between the two folds, with structural similarity abruptly shifting from one endpoint structure to the other, as seen by the TM score (Figure~\ref{fig:Speculative}:D). 

Some of the intermediates in the beam search path share a significant degree of sequence identity with both of the endpoints at the same time, resulting in states that share about 40\% sequence identity with one endpoint, but acquire the fold of the other endpoint (Figure~\ref{fig:Speculative}:C). Some of these chimeric sequences around the periphery of the transition could turn out to be metamorphic -- states that can simultaneously populate multiple structural conformations. While folding models like AlphaFold are known to assign a single conformation to metamorphic sequences with a moderate to high prediction confidence~\citep{chakravarty2022alphafold2}, experimental studies have shown that chimeric sequences of this sort can at times populate multiple conformational states~\citep{he2012mutational}.

\section{Discussion}

Here we introduce a computational technique to generate viable paths between protein sequences using stepwise single residue mutations. This method employs a greedy search protocol powered by language model embeddings to navigate through the mutational space separating the endpoint sequences. Unlike previous approaches that have used family-specific models for this task~\citep{tian2020exploring, detlefsen2022learning, mauri2023mutational}, we show that a protein language model like ESM2 can be used as~is to interpolate between arbitrary homologous protein pairs. This is enabled by the model's ability to internalize the grammar of the language of proteins~\citep{verkuil2022language}, which in turn allows it to sample states from the \textit{in-between} parts of protein morphospace. We use this capability to interpolate within and across the confines of established protein families, extending well into speculative homology.   

We find that paths between outright homologs maintain high fitness scores throughout the traversal, but those between structurally dissimilar pairs invariably pass through low fitness intermediates. There may be several possible reasons for this phenomenon. It could be the case that the search protocol is simply inadequate to tackle the more challenging problem of interpolating between two distinct structural conformations. In this scenario, paths that maintain a low value of ESM2 OFS pseudo-perplexity all the way through exist, but the beam search strategy is too crude to find them. Alternatively, it could be that chimeric sequences that lie in between two folds may increase the uncertainty in the model's predictions, leading to states that score low on fitness regardless of their true underlying viability. Another possible explanation relates to the fact that structural changes in a protein may necessitate passage through relatively unstable intermediates~\citep{bryan2010proteins}. If the model truly perceives protein stability, then it may not be possible to generate paths that maintain a low-enough value of the fitness metric in such traversals.

More importantly however, evolutionary processes can entail more than just single residue changes -- discontinuous jumps in the form of larger scale indels can significantly alter the course of evolutionary trajectories~\citep{prakash2015domain, alvarez2022creative}. Such changes, as are seen in the case of domain shuffling or fusion events~\citep{kawashima2009domain, yan2014evolutionary}, can birth altogether new protein folds as has been proposed in creative destruction~\citep{alvarez2022creative}. Novel domains that are created in this fashion need not be connected to other domains through single residue changes. Interpolating from such sequences may necessitate passing through intermediates with a low predicted fitness.

Our approach has several caveats that merit consideration. ESM2 OFS pseudoperplexity is an imperfect measure of protein fitness. Scoring metrics based on language model likelihoods, are shaped by the data distribution used to train the model~\citep{ding2024protein, shaw2023removing}. This means that the interpolation scheme may be biased to generate intermediates that tend to be similar to proteins belonging to favored clades of the model. Furthermore, while this measure has been benchmarked~\citep{kantroo2024pseudo} for comparisons between sequences that share a high degree of similarity, as seen in DMS assays~\citep{notin2024proteingym}, it may not work as a global metric to directly compare the viability of distantly related sequences. Indeed, a fitness comparison may not even be sensible for proteins that occupy different functional niches. The possible stability-dependence of the fitness metric further complicates the assessment of true functional plausibility, since somewhat unstable sequences could still be functional. Finally, we note that the known inaccuracy in the predictions of such fitness measures~\citep{notin2024proteingym} makes it likely that the paths that we find are not truly viable along their entire length. However, even though the learned fitness landscape may not be a perfect representation of the ground truth, it may still be used to gather qualitative trends about the traversal dynamics in the neutral regime. 

Our results point towards a way of inferring evolutionary relationships between proteins for which both sequence and structure are too far diverged to be informative. Distant homology inference has primarily been viewed through the lens of feature alignment. Even though the set of features being used to draw the alignment-based comparisons~\citep{finn2011hmmer, soding2005hhpred, hamamsy2024protein, llinares2023deep, johnson2024sensitive} has evolved from sequences to encompass structures and language model representations, the core of the approach remains unchanged -- identification of conserved motifs. Quantifying the ease of interpolating between two sequences presents an alternative to this approach. It may be used to measure the feasibility of an evolutionary transition mediated by single residue mutations: longer paths that are composed of more mutational steps, and those that pass through low fitness states should be harder to traverse. While pseudo-perplexity itself may not be the ideal fitness metric for this use case, the framework can accommodate other measures that better encapsulate functional plausibility. Refinements to the interpolation scheme that optimize it for both the strategy of the gameplay, along with the speed of inference can enable the framework to be used at scale. This capability can facilitate exploring the sequence-structure landscape across diverse protein families, potentially yielding new insights into the mechanics of protein evolution.

\section{Acknowledgements}

We thank Michael Abbott and Casey Dunn for close reads and useful feedback on the manuscript. This work was supported by NIH R35GM138341 (BBM). We also thank the Yale Center for Research Computing, ITS Cloud, and AWS for guidance and use of AWS Cloud computing infrastructure through Cloud Credits for Research Program.

\bibliographystyle{ieeetr}
\bibliography{deep_seq_interp.bib}

\end{document}